\documentstyle[12pt,epsf]{article}
\def\beq{\begin{eqnarray}}
\def\eeq{\end{eqnarray}}
\def\be{\begin{equation}}
\def\ee{\end{equation}}
\def\bra{\langle}
\def\ket{\rangle}

\def\half{\frac{1}{2}}    \def\thalf{{\textstyle \frac{1}{2}}}

\def\SU{\mbox{SU}}

               \def\gA{g_A}
\def\gAs{g_A^{(0)}}

        \def\calO{{\cal O}}

        \def\calV{{\cal V}}
\def\calJ{{\cal J}}        \def\calW{{\cal W}}

\def\uup{{u \uparrow}}
\def\udown{{u \downarrow}}
\def\dup{{d \uparrow}}
\def\ddown{{d \downarrow}}

\begin{document}
\baselineskip=20pt

\hfill
\parbox{4cm}{
FAU-TP3-95/6\\
hep-ph/9506342
}\\

\begin{center}
{\large  The large-$N_c$ limit and the behavior of $\gAs$ and $\gA$}
\end{center}

\begin{center}

Atsushi Hosaka$^{(1)}$ and Niels R. Walet$^{(2)}$

\vspace*{0.5cm}

$^{(1)}${\it Numazu College of Technology,
3600 Ooka, Numazu, Shizuoka, 410  Japan}

\vspace*{0.2cm}

$^{(2)}${\it Institut f\"ur theoretische Physik III,
Universit\"at Erlangen-N\"urnberg,\\ D-91058 Erlangen, Germany}

\end{center}

\vspace*{1cm}

\begin{abstract}
We investigate the isoscalar and isovector components of the axial vector
coupling constants, $\gAs$ and $\gA$
using algebraic models that lead to the correct symmetries of large-$N_c$
QCD.
Results obtained previously in various chiral models
are interpreted from this algebraic point of view.
The results of the Skyrme model and the valence quark model are
explained by simple realizations of the algebra.
\end{abstract}

\vspace*{1cm}

%=======================================================================
\section{Introduction}
%=======================================================================

In recent studies of large-$N_c$ baryons and
mesons~\cite{DaMa,Jenk,DJM,Geor}, many results derived
originally in the Skyrme model, the chiral bag model and the
chiral quark soliton model have been obtained using
algebraic methods, where the algebra is inferred from the behavior of
large-$N_c$ QCD~\cite{GerSa}.
The algebraic method does not
depend on the details of dynamics, and can provide a
clear understanding of the results obtained in specific models.
Furthermore, it provides a method for systematic calculation of higher order
terms in the $1/N_c$ expansion for meson-baryon
coupling constants, magnetic moments and other
quantities~\cite{DaMa,Jenk,DJM,Geor}.

A different but related algebraic method was developed by Amado and
collaborators
for finite $N_c$ corrections to the Skyrme model,
borrowing ideas from the interacting boson model in nuclear
physics~\cite{OBBA,ABO}.
{}From the perspective of the modern work on large-$N_c$ QCD it has now become
clear that the group considered by them is the minimal one consistent with
the large-$N_c$ behavior of QCD. This immediately raises the question whether
other realizations of the algebra exist, that have the same limit for
$N_c\rightarrow\infty$,
but lead to different predictions for finite $N_c$.
In this paper, we discuss this question looking at
the nucleon matrix elements of
$\sigma_i$ and $\sigma_i \tau^a$, which are related to the axial vector
coupling constants $\gA^{(0)}$ and $\gA$.

Our motivation is twofold:\\
(1) The fact that
those quantities were calculated successfully in
several different chiral models~\cite{HoTo,WaWa,Goek}
strongly suggests that there should exist a feature common to
those models, which is based on the underlying algebraic structure.
Our main purpose is therefore to construct an
algebraic model which is able to explain those results.\\
(2) In the algebraic method,
Skyrme model results have been interpreted to be the limit of
$N_c \to \infty$~\cite{Bard,Mano}.
Specifically, this has been explicitly shown using the so called quark
representation, the symmetric representation with Young tableau
$[N_c]$ of the SU(4) group generated by spin and isospin.
A seeming advantage of this representation is that it
produces not only the large-$N_c$ limit correctly but also
the quark model results at finite $N_c$.
It turns out, however, that this is not always the case.
For a quantity such as the nucleon
spin $\gAs$ the Skyrme model result, which is
essentially zero~\cite{BrEll}, is not reproduced in the quark representation.

Therefore,  we construct an appropriate parameterization
which reproduces both the Skyrme and the quark model limits for $\gAs$ and
$\gA$. This requirement leads to a study of models with a
dynamical symmetry group that is large enough to have $\SU(4)$ as a subgroup.
In order to get interesting models we let ourselves be guided
by some aspects of effective models of QCD, that might be particularly
relevant for the physical questions considered here. Of course
a dynamical symmetry requires that we associate some dynamics with
the models we consider. In the present note we shall concentrate
on the states, and will not discuss what Hamiltonians lead to such
a state. Fortunately, such Hamiltonians do exist. They will
be the subject of a separate study.

%=======================================================================
\section{Algebra for large-$N_c$ QCD}
%=======================================================================

In the following two sections,
we repeat some of the essential details of the large-$N_c$ algebra
as first set out in the  paper by Gervais and Sakita~\cite{GerSa}.
For that purpose  consider the pion-nucleon scattering amplitude.
To leading order, the relevant Born terms are given by
\beq
\label{TpiN}
{\cal M} \sim N_c \sum_Y \left\{
\frac{\calV^\beta_{XY} \calV^\alpha_{YZ}}{\omega + M_Z - M_Y}
- \frac{\calV^\alpha_{XY} \calV^\beta_{YZ}}{\omega - M_X + M_Y}
\right\} \, ,
\eeq
Here
$\sqrt{N_c} \calV^\alpha_{XY}$ are the Yukawa vertices with
the $\sqrt{N_c}$ dependence factored out,
and $M_{X \cdots}$ are the masses of the baryons.
In these quantities, subscripts
$X$, $Y$, $\cdots$ specify the quantum numbers for
the baryons which are in the fundamental representations of the spin and
isospin symmetries,
$X  \equiv  (J(\mbox{spin}), I(\mbox{isospin}))$
etc, while $\alpha$, $\beta$ stand for the quantum numbers  of the pions.
The latter label the adjoint
representations of the same spin-isospin group, since the pion
carries isospin one and it couples with the nucleon through the
P-wave.
{}From unitarity and also from the Witten's large-$N_c$ counting
rule~\cite{Witt79},
the scattering amplitude is bounded from below
for arbitrary $N_c$.
Thus
the two terms in (\ref{TpiN}) must cancel each other in the limit
$N_c \to \infty$.
This is the primary consistency condition for large-$N_c$
QCD~\cite{DaMa,DJM}. It
is satisfied if the mass differences of the tower of states built
upon the nucleon go to zero in this limit
and the following relation holds,
\beq
\label{Tcommute}
\sum_Y \left(
\calV^\beta_{XY} \calV^\alpha_{YZ} - \calV^\alpha_{XY} \calV^\beta_{YZ}
\right) \stackrel{N_c \to \infty}{\longrightarrow} 0 \, .
\eeq
Here the left hand side must be suppressed as proportional to
$1/N_c$.
Equation (\ref{Tcommute})
is now regarded as a matrix equation for $\calV^\alpha$ whose components are
$\calV^\alpha_{XY}$.

In the limit $N_c \to \infty$,
combining the spin and isospin algebra for light flavors,
$K = \SU(2) \times \SU(2)$, with that generated by (\ref{Tcommute}),
we obtain the following commutation relations:
\beq
\left[ \calJ^a, \calJ^b \right] &=& i f^{abc} \calJ^c \, , \nonumber \\
\label{KTalge}
\left[ \calJ^a, \calV^\alpha \right] &=& i D(a)^\alpha_\beta \calV^\beta \, ,
\\
\left[ \calV^\alpha, \calV^\beta \right] &=& 0 \nonumber \, .
\eeq
Here $\calJ^a \in K$ and $f^{abc}$ are the generators and
the structure constants for the group $K$.
$D(a)^\alpha_\beta$ are the matrices of the generators  $\calJ^a$
in the basis of $\alpha$:
$D(a)^\alpha_\beta \equiv \bra \alpha|J^a|\beta\ket$.
Equations (\ref{KTalge}) shows that $\calJ^a$ and
$\calV^\alpha$ form the non-compact group algebra
$G = K \times T$, the semi-direct product
of $K$ by the Abelian group $T$ generated by $\calV^\alpha$.
Because of this non-compact nature of the group,
the baryon states form an infinite tower of states with spin
equal to isospin
$(I,J) = (\half,\half), (\frac{3}{2},\frac{3}{2})$, etc.

%===================================================================
\section{Representations of the large-$N_c$ group}
%===================================================================
In Ref.~\cite{GerSa} the construction of representations for
the large-$N_c$ group $G = K \times T$
is performed  by
the method of induced representations.
Equivalently one can use the technique
of group contractions
starting from the compact group  $\SU(4)$.
Here we see naturally how the extrapolation to  infinite $N_c$  is performed.

The first step in the argument is based on
the fact that the number of generators
of $K\times T$, with
$K = \SU(p) \times \SU(q)$ and the $T$ a direct product of
the adjoint representations of $\SU(p)$ and $\SU(q)$,
\beq
N=(p^2-1) + (q^2 - 1) + (p^2-1)(q^2-1) = (pq)^2 - 1 \, ,
\eeq
is the same as the number of generators of the $\SU(pq)$ group.
Let us look at the  $\SU(p) \times \SU(q)$ subgroup of
$\SU(pq)$. Let us denote the generators of the subgroup by $\calJ$,
and the remaining generators by $\calW$. The algebra now
takes the form
\beq
\left[ \calJ^a, \calJ^b \right] &=& i f^{abc} \calJ^c \, , \nonumber \\
\label{SUalge}
\left[ \calJ^a, \calW^\alpha \right] &=& i D(a) ^\alpha_\beta \calW^\beta
\, , \\
\left[ \calW^\alpha, \calW^\beta \right] &=&
i \left( g_1^{\alpha \beta \gamma} \calW^\gamma
+ g_2^{\alpha \beta c} \calJ^c \right) \, . \nonumber
\eeq
Here $f$, $g_1$, $g_2$ are the structure constants.
For the case of $p = q = 2$, the constants $g_1$ vanishes and
the last commutators of (\ref{SUalge}) can be
written as
\beq
\label{WWcommu}
[ \calW^{ia}, \calW^{jb}] =
\frac{i}{4} \epsilon_{ijk} \delta_{ab} J^k
+ \frac{i}{4} \epsilon_{abc} \delta_{ij} I^c \, ,
\eeq
where $J^i$ and $I^a$ are the generators of the spin and isospin group.

Consider now the $N$-dimensional symmetric
representation of $\SU(4)$ and suppose that the
matrix elements of $\calW$ are of order $N$.
The
group contraction $\SU(4) \to G=K\times T$ can be performed by first replacing
$\calW^\alpha$ by
$\calV^\alpha \equiv \calW^\alpha/N$.  In the limit $N \to \infty$
the algebra (\ref{SUalge}) written in terms of $\calV$ then reduces to
the algebra (\ref{KTalge}).
The index $N$ can then be identified with the number of colors in what has been
called the ``quark representation'' (which refers to valence quarks)
as shown in the following.

Consider the wave function of a baryon. It can be written as a
direct product of orbital, spin, isospin (flavor) and color parts.
Since the color part is totally antisymmetric (it has Young tableau
$[(1)^{N_c} ]$)
the rest of the wave function must be totally symmetric, with
Young tableau $[N_c]$.
For ground state baryons,  quarks are assumed to be
in the lowest S state, which gives a symmetric orbital wave function, and thus
the spin and isospin part must be symmetric as well.
The symmetric representation of $\SU(4) \supset \SU(2) \times \SU(2)$
is specified by an index $N$, which must now be identified with the number of
colors $N_c$.

Let us introduce operators
which generate the fundamental representations
for spin and isospin group,
$\alpha_{\mu \nu}$.
For instance, $\alpha_{\udown}$ annihilates a spin down up quark.
The symmetric $N_c$ representation of the $\SU(4)$ group is
then generated by the defensive hedgehog state~\cite{OBBA,ABO}
(the name defensive hedgehog state derives from the Skyrme model,
where this state corresponds to a pion field that point radially outwards,
as on a hedgehog defending itself)
\beq
|N_c;h_q\ket &=& \sqrt{\frac{1}{N_c !}}\left[
\frac{1}{\sqrt{2}}
(\alpha^\dagger_{\udown}-\alpha^\dagger_{\dup}) \right]^{N_c} |0\ket \, ,
\label{Nquark}
\eeq
where on the left hand side the subscript $q$ indicates
that the state is the quark representation. The operator in square brackets
is exactly what is obtained by coupling the spin and isospin to zero.
This coupling to ``grand-spin'' $K=0$ is typical of a hedgehog state.
The state (\ref{Nquark}) breaks rotational and iso-rotational symmetries.
This is possible since all states obtained by applying an isospin
rotation operator $R(A)$ (where $A$ is the two-by-two unitary matrix
specifying the transformation) on the
defensive hedgehog state (\ref{Nquark}) are degenerate,
\beq
|N_c; h_q[A]\ket &\equiv& R(A)|N_c;h\ket  \nonumber \\
\label{NhA}
& & \hspace*{-2cm} = \;
\sqrt{\frac{1}{N_c !}}
\left[ \frac{1}{\sqrt{2}}
\left( D^{1/2}_{uu}(A) \alpha^\dagger_{\udown}
+ D^{1/2}_{du}(A) \alpha^\dagger_{\ddown}
- D^{1/2}_{dd}(A) \alpha^\dagger_{\dup}
- D^{1/2}_{ud}(A) \alpha^\dagger_{\uup} \right) \right]^{N_c} |0\ket\, .
\eeq
Here $D^{1/2}_{\mu \nu}(A)$ are the $D$-functions of rank $1/2$.
The semi-classical nature in the limit $N_c \to \infty$ is reflected
in the overlap function,
\be
\bra N_c;h[A^\prime]| N_c; h[A]\ket =
\left[ \cos \theta_{AA^{\prime\dagger}} \right]^{N_c} \, ,
\ee
where $\theta_{AA^{\prime\dagger}}$ represents symbolically the relative
angle between the ``orientations'' $A$ and $A^\prime$.
After an appropriate rescaling of the states this overlap goes to
a delta function~\cite{WaletHosaka}
\be
\bra N_c;h[A^\prime]| N_c; h[A]\ket =
\delta(A-A^\prime) .
\ee
This sharp peaking is characteristic of a semiclassical limit.

%===================================================================
\section{Nucleon matrix elements}
%===================================================================

The nucleon state $|N\ket$ can be projected out from the hedgehog state
$|h\ket$ by taking the appropriate combination of rotated hedgehogs
\cite{GerSa,Mano},
\beq
\label{project}
|N\ket \propto \int d[A] D^{1/2*}_{t,-m}(A)\,  R(A) \, | h\ket \, .
\eeq
Here the nucleon third components of spin and isospin equal $m$ and $t$.
The nucleon matrix elements of an observable $\calO_N$ can now be computed
as
\beq
\calO_N = \frac{\bra N |\calO |N\ket}{\bra N | N\ket} \, ,
\eeq
where the denominator is needed for normalization.

Let us now calculate the two axial-vector
coupling constants for the nucleon, $\gA^{(0)}$ and $\gA$.
For $\gA^{(0)}$ we should consider the ${\rm U}(1)_A$
effects seriously,
but these effects will be ignored here, and we shall see how a simple
construction behaves for  these quantities.

The actual computation is performed most conveniently using the Euler angles
$\alpha$, $\beta$ and $\gamma$ for the rotation matrix $A$.
The computational procedure is straightforward \cite{WaletHosaka}
and we just give the results here.
$\gAs$ and $\gA$ are defined by the nucleon matrix elements of
${\displaystyle S_3 \equiv \sum_{n=1,N_c} \sigma_i(n)}$ and
${\displaystyle T_{33} \equiv \sum_{n=1,N_c} \sigma_i(n) \tau_a(n)}$.
The expectation values in the $p\downarrow$ state are
\beq
\bra p \downarrow | S_3 | p \downarrow \ket
&=&  - N_c \frac{ {\displaystyle \int d[A] \,
C_\beta ^2 \, S^2 \, C^{N_c-1} } }
{{ \displaystyle \int d[A] \, C^{N_c+1} }} \nonumber \\
\label{Nsigma}
&=&  - \frac{N_c}{N_c} \; = \; -1 \, .
\eeq
and
\beq
\bra p \downarrow | T_{33} | p \downarrow \ket
&=& - \frac{N_c}{3} \frac{ {\displaystyle \int d[A]
\left( C^{N_c+1} + C^{N_c-1} S^2 \right) }}
{{\displaystyle \int d[A] \; C^{N_c+1} }} \nonumber \\
\label{Nsigmatau}
&=& - \frac{N_c}{3} \left( 1 + \frac{2}{N_c} \right) \, .
\eeq
Here we have introduced the notations
$
C_\beta = \cos \frac{\beta}{2} \, , \;
C = \cos \frac{\beta}{2} \, \cos\frac{\alpha + \gamma}{2} \, , \;
S = \sin \frac{\alpha+\gamma}{2}
$, and
$d[A] = \sin \beta \, d\beta \, d\alpha \, d\gamma$.
The minus signs appear because the matrix elements
are evaluated for the $p \downarrow$ state.
The result (\ref{Nsigma}),
$\bra p\downarrow|S_3|p\downarrow\ket = -1$, as
independent of $N_c$ is trivial,
since the nucleon spin here is entirely carried
by the intrinsic quark spin, which is the result of the
non-relativistic valence quark model.
On the other hand, the result (\ref{Nsigmatau}) depends on $N_c$;
it becomes (neglecting the
minus sign) 5/3 when $N_c$ = 3 as in the valence quark model,
while it approaches $N_c$ in the limit $N_c \to \infty$ as
corresponding to the Skyrme model result.
Thus the quark representation is believed to
interpolate between the quark model and the Skyrme model
results when $N_c$ goes from 3 to $\infty$~\cite{Mano}.
This is the case for $\gA$, but is not true for the other
matrix element of the nucleon spin,
$\gA^{(0)} \sim \bra p \downarrow | S_3 | p \downarrow \ket$.

%===================================================================
\section{Other algebraic realizations}
%===================================================================

In this section we consider other algebraic realizations for the group
$\SU(4)$ which is contracted to the large-$N_c$ algebra
in the limit $N_c \to \infty$.
By doing this, we will see a novel behavior of the matrix elements
for $\gAs$ and $\gA$.

First we consider the relativistic effects  which involve the lower components
of the wave function
with  orbital angular momentum $L$ = 1.
Formally this can be achieved by first extending the rotational group SU(2)
to $\SU(2)_S \times {\rm O}(3)_L$ for spin ($S$) and orbital angular
momentum ($L$)
group, and then take its diagonal subgroup:
$\SU(2)_S \times {\rm O}(3)_L \supset \SU(2)_{J = S + L}$,
where the generators of the diagonal group is the sum
$\vec J = \vec S + \vec L$.
The $\SU(2)_J$ group is now combined with the isospin $\SU(2)_I$ group
to form the desired chain $\SU(4)_q \supset \SU(2)_J \times \SU(2)_I$, where
the subscript $q$ on the left hand side indicates
that this algebra is realized by the quarks.
We shall not discuss
other possible dynamical origins of orbital excitations in this note, which
are discussed in great detail in Ref.~\cite{WaletHosaka}.

The second extension is to include explicit pions, that also form
a hedgehog state. Since a pion can form a  grand-spin $K = 0$ state
through the coupling of its unit
isospin and $L = 1$ orbital angular momentum (P-wave),
the relevant group is ${\rm O}(3)_I \times {\rm O}(3)_L$,
which we once more imbed in an $\SU(4)$ group.
Therefore, the extended group for the hedgehog quarks and the pions would
be
\be
\SU(4)_q \times \left[ {\rm O}(3)_I \times {\rm O}(3)_L \right]_\pi
\subset \SU(4)_q \times \SU(4)_\pi.
\ee
Once again, we pick up the diagonal subgroup,
$
\SU(4)_q \times \SU(4)_\pi \supset \SU(4)_{q+\pi}.
$

The relativistic hedgehog quarks are now written as
\beq
\label{hqrel}
|N_c;h\ket =
\left(
\begin{array}{c}
c_1 \left[ [\thalf, \, 0]^{J=1/2},\, \thalf\right]^{K=0} \\
c_2 \left[ [\thalf, \, 1]^{J=1/2},\, \thalf\right]^{K=0}
\end{array}
\right)^{N_c} |0\ket
\equiv
\left[ \left(
\begin{array}{c}
c_1  \\
c_2 \vec \sigma \cdot \hat x
\end{array}
\right) \chi^\dagger \right]^{N_c} |0\ket\, .
\eeq
Here the coupling scheme is $[[S,L]^J,\, I]^K$ and
$\chi^\dagger \equiv \left[ [\thalf, \, 0]^{J=1/2},\, \thalf\right]^{K=0}
= \frac{1}{\sqrt{2}}
(\alpha^\dagger_{\udown}-\alpha^\dagger_{\dup})$.
The coefficients $c_1$ and $c_2$, $|c_1|^2 + |c_2|^2 = 1$,
 dictate the ratio of the upper ($L = 0$) and
lower ($L = 1$) components, the precise values of which are
determined dynamically.
The last expression of (\ref{hqrel}) is useful in actual computation.
We wrote the matrix $\sigma$ instead of the algebraic operator $S$
to exhibit the fact that it only acts on a {\em single} operator $\chi$.
The transformation $\vec\sigma\cdot\hat x$ acts on the spin components
of the operator $\chi$.

The matrix elements for $S_3$ and $T_{33}$ are computed in a
straightforward manner and the results are
\beq
\bra p \downarrow | S_3 |  p \downarrow\ket
= - \left( |c_1|^2 - \frac{1}{3} |c_2|^2 \right)
\equiv - \gAs
\eeq
and
\beq
\bra p \downarrow | T_{33} |  p \downarrow\ket
= - \left( |c_1|^2 - \frac{1}{3} |c_2|^2 \right)
\frac{N_c}{3}\left( 1 + \frac{2}{N_c}\right)
\equiv - \gA \, .
\eeq
We find the same suppression factor
$\left( |c_1|^2 - \frac{1}{3} |c_2|^2 \right)$ for both the
 matrix elements.
They have been included in relativistic quark
model calculations such as in the bag  and the chiral quark soliton
models~\cite{HoTo,WaWa,Goek}.
For the maximally relativistic case of massless quarks, it becomes
0.654  as in the MIT bag model.
Accordingly,  $\gAs=0.654$ and $\gA = 5/3 \cdot 0.654$ = 1.09~\cite{MIT}.
If we wish to reproduce the experimental value of $\gAs$
we need a suppression factor of about 1/4.
This implies an unrealistically small value for
$\gA \sim 5/3 \cdot 1/4 \sim 0.4$.
The inclusion of the asymptotic
one pion contribution does not help very much
(See Fig.~1, as well as the discussions below).
It increases $\gA$ about by 50 \% \cite{Jaffe},
but still $\gA$ is only of the order of  0.6.
The Skyrme model result $\gAs = 0$ is obtained when the
suppression factor vanishes but in this case $\gA$ is also zero,
which is inconsistent with the Skyrme model.

\begin{figure}[htb]
\epsfxsize=15cm
\centerline{\epsffile{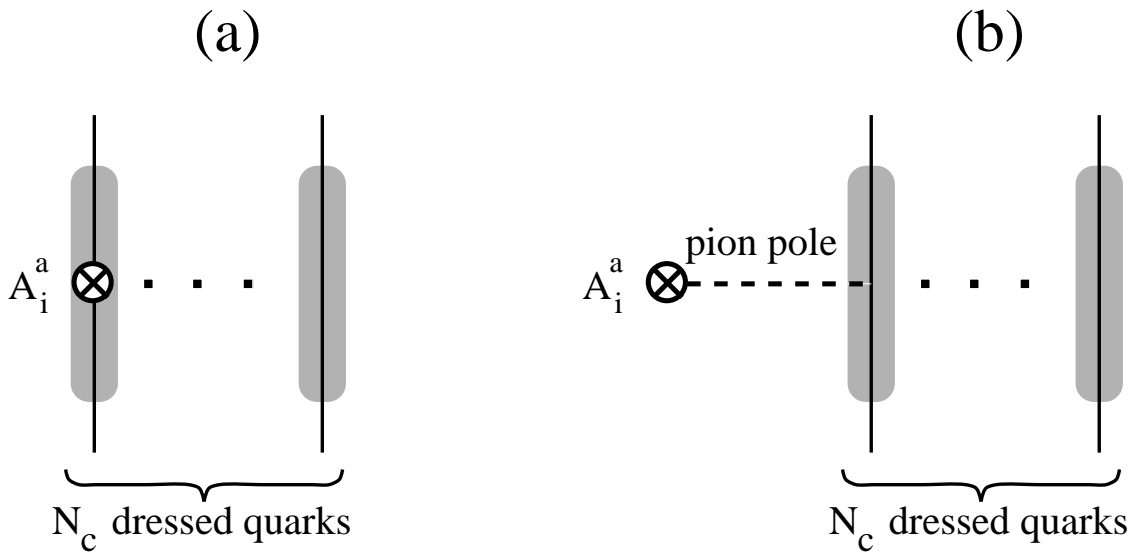}}
\caption{
Two components of the isovector axial vector current $A^a_i$: (a) the
current which directly couples to the quark line, and (b) the current
which couples to the pion pole term.  In both cases, the quarks are
dressed by pion clouds as indicated by the hatched area.  The number of
the pions in the cloud around the quarks is related to $N_\pi$ in
(\protect{\ref{qandh}}). }
\end{figure}

Next we consider the pionic effect on the $N_c$ quark state.
This must be distinguished from the asymptotic
one pion contribution from the tail
region but is rather
related to the finite range effect which generates
the hedgehog structure through non-linear interactions
in the large-$N_c$ limit (Fig.~1).
We consider an ansatz of the direct product of the $N_c$ quarks and
$N_\pi$ pions. The $\SU_\pi(4)$ algebra can be realized in terms of
the same type of operators as the quark algebra. Since a
pion carries spin one, the hedgehog form
can be expressed as
$\left[
\frac{1}{\sqrt{2}}
(\alpha^\dagger_{\udown}-\alpha^\dagger_{\dup}) \right]^2$.
It is convenient to lump the
quarks and pions together \cite{WaletHosaka} and write
\beq
\label{qandh}
|N_c;h_{q + \pi} \ket =
\sqrt{\frac{1}{(N_c+2N_\pi)!}}\left[
\frac{1}{\sqrt{2}}
(\alpha^\dagger_{\udown}-\alpha^\dagger_{\dup}) \right]^{N_c + 2N_\pi}
|0\ket\, .
\eeq
The meaning of the number of the pions is not very clear yet, but
is related to
how the nucleon spin is partitioned between quarks and pions.
In a large-$N_c$ baryon where the bound state of $N_c$ quarks is treated
in the Hartree approximation, the number of the pions in the baryon would be
expected to be proportional to $N_c$, since in this approximation
it is $N_c$ times the number of the pions around a single quark, which
one expects to be of order one.
In the following argument, however, $N_c$ and $N_\pi$ are treated as
independent parameters in order to understand the results in what follows.

Now the nucleon matrix elements for $S_i$ and
$T_{ia}$ are computed in the same manner as before.
Note that those operators act directly on the $N_c$
quarks as illustrated in Fig.~1(a).
The results are
\beq
\label{Sqpi}
\bra p \downarrow | S_3 | p \downarrow \ket &=& - \frac{N_c}{N_c + 2 N_\pi}
\equiv - \gAs \, , \\
\label{Tqpi}
\bra p \downarrow | T_{33} | p \downarrow \ket &=&
- \frac{N_c}{3} \left( 1 + \frac{2}{N_c + 2 N_\pi} \right) \equiv - \gA \, .
\eeq
These results have interesting implications.
The nucleon spin $\gAs$ becomes less than unity for a
finite number of pions $N_\pi \neq 0$, where
a part of the nucleon spin is carried by the angular momentum of pions.
When $N_\pi=0$, Eq.~(\ref{Sqpi}) reduces
to the quark model result, where the entire
nucleon spin is carried by the quark spin,  and when
$N_\pi \to \infty$ we obtain the skyrmion result, where the nucleon spin is
entirely carried by the pion cloud~\cite{Waka}.
The same thing holds also for the second term of $\gA$ in (\ref{Tqpi}).
In particular, the realization of the Skyrme model results in the limit
$N_\pi \to \infty$ as shown here,
rather than $N_c \to \infty$, is interesting.
This result may be understood  by interpreting
the Skyrme soliton as a coherent superposition
of infinitely many pions.
At this point, it is interesting to recall the recent result by
Dorey and Mattis, who have explicitly shown that the Skyrmion is an
ultraviolet fixed point of a chiral bag model, where the
coupling between the pion and the bare nucleon disappears~\cite{DorMatt}.
This implies that the dynamics of the meson-baryon system is completely
described by the pion alone.

Let us now try to make a
rough estimate for $\gAs$ and $\gA$.
In the physical nucleon, both effects,
the relativity through the $L=1$ component and the pionic effects,
must be considered. Some algebra shows that this can be done by
multiplying both (\ref{Sqpi}) and (\ref{Tqpi}) by the reduction factor
$|c_1|^2 - \frac{1}{3} |c_2|^2$. A reasonable  estimate for this quantity
is $\sim 0.7$.
For $N_c = 3$, $\gA$ of (\ref{Tqpi}) becomes
\beq
\label{gA2}
\gA \sim 0.7 \left(1 + \frac{2}{3+N_\pi} \right) \, .
\eeq
The pion contribution as depicted in Fig.~1(b) must still be added to
this result.
In the chiral limit this can be estimated to be about 50 \% of the quark
contribution of (\ref{gA2})~\cite{Jaffe}, and so we get for the total $\gA$
\beq
\label{gAtot}
\gA &\sim& 1.5 \times 0.7 \left(1 + \frac{2}{3+N_\pi} \right) \nonumber
\sim \left(1 + \frac{2}{3+2N_\pi} \right)  \, .
\eeq
The experimental value of $\gA \sim 1.3$ is then reproduced when
$N_\pi \sim 2$.
Note that the $1/N_c$
correction term in (\ref{gAtot}) is about 30 \%, which
is in good agreement with the previously quoted numbers in the chiral
quark soliton model and in the chiral bag model~\cite{HoTo,WaWa}.
Using the same parameters, the nucleon spin $\gA^{(0)}$ comes out to be
\beq
\gAs \sim 0.7 \frac{3}{3 + 2 N_\pi} \sim 0.3  \nonumber \, .
\eeq
It is indeed remarkable that such a simple algebraic
method can be used to describe both $\gA$ and $\gA^{(0)}$ simultaneously
in good agreement with experiments.

\section{Summary}

In this note
we have investigated algebraic models that contain a subgroup
$\SU(4)$ whose contractions reduce to the large-$N_c$ algebra for QCD.
We have explicitly constructed a realization which interpolates the
skyrmion and the quark model results for $\gAs$ and $\gA$.
In order to find such a result we were
inspired by models to include explicit pionic degrees of freedom.
The agreement of the present calculations with experiments and with previous
model calculations suggests that those quantities are
strongly governed by the underlying algebraic structure, rather than
their detailed dynamical details, as also
implied by the large-$N_c$ limit of QCD.

A detailed discussion of several
algebraic models as phenomenological models compatible with the
large-$N_c$ behavior of QCD will be presented in Ref.~\cite{WaletHosaka}.

\vspace*{0.5cm}

\noindent
{\bf Acknowledgments}

\noindent
A.H. acknowledges the Institute for Nuclear Theory at the University of
Washington for its hospitality and the Department
of Energy for partial support during the completion of this work.
The research of N.R.W. is supported in part by the German Federal
Minister for Research and Technology (BMFT).
\\

\end{document}